\documentclass[twocolumn,aps,showpacs,prl,epsf,graphics,psfig]{revtex4}
\usepackage{graphicx}

\begin{document}
\title{Multiscale QM/MM Molecular Dynamics Study on the First Steps of Guanine-Damage by Free Hydroxyl Radicals
in Solution}

\author{Ramin M. Abolfath$^{1,2}$, P. K. Biswas$^{3,4}$, R. Rajnarayanam$^5$, Thomas Brabec$^2$,
Reinhard Kodym$^6$, Lech Papiez$^7$}

\affiliation{
$^1$School of Natural Sciences and Mathematics, University of Texas at Dallas, Richardson, TX 75080 \\
$^2$Physics Department, University of Ottawa, Ottawa, ON, K1N 6N5, Canada \\
$^3$Department of Physics, Tougaloo College, Tougaloo, MS 39174 \\
$^4$Laboratory of Computational Biology, NHLBI, National Institutes of Health, Rockville, MD 20894 \\
$^5$Department of Pharmacology, University of Buffalo, Buffalo, NY 14260 \\
$^6$Department of Radiation Oncology, University of Texas, Southwestern Medical Center, Dallas, TX 75390\\
$^7$Indiana University Cyclotron Facility, Bloomington, IN 47408\\
}

\date{\today}

\begin{abstract}
Understanding the damage of DNA bases from hydrogen abstraction by free OH radicals is of particular importance to reveal the effect of hydroxyl radicals produced by the secondary effect of radiation. Previous studies address the problem with truncated DNA bases as ab-initio quantum simulation required to study such electronic spin dependent processes are computationally expensive. Here, for the first time, we employ a multiscale and hybrid Quantum-Mechanical-Molecular-Mechanical simulation to study the interaction of OH radicals with guanine-deoxyribose-phosphate DNA molecular unit in the presence of water where all the water molecules and the deoxyribose-phosphate fragment are treated with the simplistic classical Molecular-Mechanical scheme. Our result illustrates that the presence of water strongly alters the hydrogen-abstraction reaction as the hydrogen bonding of OH radicals with water restricts the relative orientation of the OH-radicals with respective to the the DNA base (here guanine). This results in an angular anisotropy in the chemical pathway and a lower efficiency in the hydrogen abstraction mechanisms than previously anticipated for identical system in vacuum. The method can easily be extended to single and double stranded DNA without any appreciable computational cost as these molecular units can be treated in the classical subsystem as has been demonstrated here.
\\ \\
Corresponding authors: Ramin Abolfath (ramin.abolfath@utdallas.edu), Pradip K Biswas (pbiswas@tougaloo.edu)
\end{abstract}
\maketitle

It is known that the megavoltage radiation (X/$\gamma$-rays, $\alpha$-particles, and heavy ions) ionize the water molecule and create neutral OH-radicals~\cite{EricHall:book}.
Various effects of the ionizing radiation on the biological systems that ranges from the development of genetic aberrations, carcinogenesis to aging have attributed to the role of free radicals~\cite{Aydogan2008:RR}.
In particular OH-radicals are major contributors in the single/double strand breaking of the DNA molecules as the 2/3 of the surrounding environment of the DNA molecules in the cell-nucleus is composed of water molecules.
OH-radical, a polar diatomic molecule is highly reactive.
They are electrically neutral with a magnetic moment that is produced by nine electrons with an unpaired spin in the outermost open shell.
In-vivo, they have a very short life-time that is reported within a few nano-seconds~\cite{Sies1993:EJB}.
Because of their short life-time, OH radicals generated within 1 nm from the surface of the DNA molecule can remove a hydrogen ion and form a water molecule.
The high reactivity of radical OH is attributed to the possible pairing of opposite spin electrons from OH and the target molecule. Due to reactivity to solvent accessibility, a free radical reaches a DNA active site and removes a hydrogen ion from sugar moiety (e.g., a deoxyribose)
~\cite{Balasubramanian1998:PNAS,Tullius2005:COCB,Pogozelski1998:CR} or nucleobase~\cite{Burrows1998:CR,Chatgilialoglu2011:CRT}.

The motion of the electrically neutral OH-radical is governed by the thermal diffusion process. Like any diatomic molecule, their energy can be expressed as the sum of electronic, vibrational, translational and rotational parts~\cite{Herzberg:Book}.
The conformation of the molecular levels with the environment during the dynamical motion of the molecule allows the molecule to find its optimal chemical pathway.
From the point of view of the tetrahedral conformation of the electronic charge distribution in Oxygen, the hydrogen abstraction pathways by OH are expected to show angular anisotropy with a peak around $109.5$ degrees for H-OH angle.
Recent simulations of the hydrogen abstraction in the vacuum by
OH-radical~\cite{Mundy2002:JPC,Close2008:JPC,Abolfath2009:JPC,Abolfath2011:JPC,Kumar2011:JPCB} also observe that the occurrence of the chemical reaction strongly depends on the relative orientation of the OH radical and the H atom of the donor molecule. In vacuum simulation, the radical OH is quasi free to orient itself appropriately with guanine hydrogens. However,
in solution, the OH radicals are expected to form hydrogen bonding with water molecules which would in turn restrict their relative orientations with respect to guanine-hydrogens and we may see a retardation in hydrogen abstraction. Thus, understanding the role of water molecules for quantitative analysis of the DNA initial damage by ionizing radiation in a realistic environment is necessary. However, the exponential growth of computational procedure with the system size is the main challenge in performing realistic calculations. A realistic environment contains of thousands of water molecules and is too expensive to be handled by any ab-initio calculation - which is required to describe the H-abstraction reaction. The multiscale, hybrid Quantum-Mechanical-Molecular-Mechanical (QM/MM)) method is a practical approach that allows significantly lower the computational effort, yet suitable to address the electronic rearrangement of the active site and consequent breaking and forming of chemical bonds. In a hybrid QM/MM, a small subsystem containing the active site of the reacting molecules is treated with an ab-initio QM method while the rest of the system including the solvent, if any, is treated with a classical Molecular-Mechanical (MM) force-field. Application of a QM method is essential for the calculation of the redistribution of electronic charges of the active site on-the-fly, as needed for the conformations of the dynamical trajectory to simulate the chemical reaction pathways.

In order to address the issue of hydrogen abstraction from a realistic system of a DNA guanine base~\cite{Miaskiewicz1994:JACS,Mundy2002:JPC,Close2008:JPC,Abolfath2009:JPC,Abolfath2011:JPC,Kumar2011:JPCB}, we consider a guanine base coupled with a deoxyribose and a phosphate backbone truncated at both ends of a single stranded DNA base unit - we refer it as a DNA residue which is one oxygen short (the O3' from the adjacent residue) to that of a nucleotide. During QM/MM, the deoxyribose and the phosphate group are treated in the MM subsystem and hence this truncation is not expected to have any direct consequences in the electronic distribution of the system and hence in the hydrogen abstraction process. However, it would certainly have truncation effect coming through the perturbation of the QM system by MM charges as we will always have with any truncated system. Our overall system consists of this DNA residue dissolved in water with one water molecule close to the guanine NH2 group converted in to a radical OH by deleting a hydrogen. The solvated system was first given a short energy minimization using OPLS force field~\cite{opls, opls-liq} so as to eliminate any bad contacts with water molecules arising from solvation. The OH-radical seems to form hydrogen bonding with the NH$_1$ and NH$_2$ groups of guanine. To understand the hydrogen abstraction mechanism, the system is then subjected to the hybrid GROMACS-CPMD~\cite{Biswas2005:JCP} Quantum-Mechanical-Molecular-Mechanical (QM/MM) simulation protocol where the active site consisting of the OH and the guanine base are treated in the QM subsystem and the deoxyribose, the truncated phosphate backbone and water are treated in the MM subsystem. As the deoxyribose and the truncated phosphate backbone are treated in the classical MM model, and they lie in close viscinity of the active site treated in QM we do not use end capping atoms for the phosphate unit as they will bring additional point charges close to the QM subsystem.
The QM subsystem is optimized with a plane-wave based DFT model as implemented in CPMD. The QM/MM partition is being made at the N$_9-$C bond connecting the guanine nitrogen to the deoxyribose. At the guanine-deoxyribose N-C bond-cut the link-atom prescription of QM/MM is used to saturate the QM subsystem for the QM simulation. The rest of the system - the deoxyribose, the phosphate, and the water are treated in the MM subsystem using the Molecular Mechanics force field OPLS~\cite{opls, opls-liq} as implemented in GROMACS~\cite{gromacs}.

The overall computational box size is taken as {$(42.22 \AA \times 49.14 \AA \times 40.78 \AA)$} which contains the whole system: the guanine-based DNA residue (guanine-deoxyribose-phosphate unit), an OH radical, and 2811 TIP3P~\cite{tip3p} water molecules. A subset of the system containing the DNA residue and water within about $5 \AA$ is shown in Figure 1a. For the QM subsystem, which contains the guanine and radical OH, a cubic cell of size  {$(14.82 \AA \times 18.52 \AA \times 10.58 \AA)$} is used together with Poisson solver of Martyna and Tuckerman \cite{Martyna1999:JCP} for the wave function optimization in CPMD. For both MM and QM/MM, we use a cutoff of $10 \AA$ for the Coulomb and van der Waals interactions. In GROMACS-CPMD QM/MM, CPMD provides the optimized wave function and forces for the QM subsystem which is appropriately perturbed by the MM subsystem and GROMACS controls Molecular Dynamics the simulation. For our QM/MM, we have performed a constant pressure and temperature (NPT) MD simulation with a reference temperature of 300K and a time step of 1fs. Our CPMD wave-function optimization is implemented in a plane-wave basis within local spin density approximation (LSDA) with an energy cutoff of 40 Rydberg (Ry), and with Becke \cite{Becke1988:PRA} exchange and Lee-Yang-Parr (BLYP) gradient-corrected functional \cite{Lee1988:PRB}. Norm conserving ultrasoft Vanderbilt pseudo-potentials were used for oxygen, hydrogen, nitrogen and carbon. Three different sets of simulation were performed - two in solution and one in vacuum. For guanine spin singlet ($s=0$), which corresponds to a doublet multiplicity for the guanine-OH system, two simulations were performed - one with solvent water and the other without solvent. For guanine spin triplet ($s=1$), which could lead to a quartet multiplicity for the guanine-OH system, only one simulation has been performed with solvent. This is to be mentioned here that for this latter system, an earlier simulation in vacuum \cite {Abolfath2009:JPC,Abolfath2011:JPC} performed using CPMD, found no hydrogen abstraction from guanine.
The results presented in this work are representative of several runs of QMMM molecular dynamics, starting from a large number of possible initial conditions.

Our hybrid QM/MM simulations consist of two classes of spin-restricted calculations, as the total spin along the quantum axis is subjected to the constraints $S_z=1/2$, and $3/2$, corresponding to doublet and quartet spin configurations. In both spin-state calculations, same initial configurations were used for the guanine-OH subsystem which was taken from an OPLS force-field generated energy minimized structure of the total system in the presence of solvent water.
To understand the effect of solvent on hydrogen abstraction, we perform QM/MM molecular dynamic simulations of the MM energy minimized structure of the solvated system (guanine-deoxyribose-phosphate-OH) obtained with OPLS force-field.
The initial configuration here corresponds to H-OH distance of $1.92\AA$ and H-O-H angle of $108.92$ degrees with the NH group hydrogen
The initial MM energy minimized configuration is characterized by hydrogen bonding of OH with NH and NH$_2$ groups of guanine and as well as with one solvent water. Where the H-bond lengths for H$_{N_1}-$OH,  H$_{N_2}-$OH, and H$_{sol}-$OH are given by $1.92\AA, 1.97\AA, 1.68\AA$, respectively.
The initial bond angle for H$_{N_1}-$O$-$H is $108.42$ degree and that of H$_{N_2}-$O$-$H is $153.95$ degrees. Consequently, the initial configuration, which is obtained by MM energy minimization, places H$_{N_1}$ in a favorable position to form water with the radical OH (since H$_{N_1}-$O$-$H angle is $108.42$ degree which is close to the tetrahedral angle of $109.5$ degrees).
The same initial structure has been used to study the vacuum property by simply deleting the water from the solvated energy minimized configuration. Using a similar system in vacuum with a H-OH distance of $1.5\AA$, it has been reported earlier~\cite{Abolfath2009:JPC,Abolfath2011:JPC} that the hydrogen abstraction does depend on the spin of guanine. For guanine spin singlet (guanine-OH doublet), the hydrogen abstraction is found to start in about 0.07 ps in vacuum and in 0.1 ps in solution. For a time step of 2 fs, the simulation exhibits numerical instability with the abstracted hydrogen is found to have strong kinetic energies leading to bouncing back to guanine and then lost to water. As the time step is reduced to 1 fs, numerical stability returns back and we notice a gradual process of hydrogen abstraction from guanine. Even with 1 fs time step, the kinetic energy of the abstracted hydrogen remains high. For simulation of the system in vacuum, this KE is found to get translated into the KE of the of the newly formed water and the resulting water molecule is found to move away quickly from the guanine. However, in the presence of solvent, the translational motion of OH is being restricted by its hydrogen bonding with solvent water molecules and thus we see quite a lot of  fluctuation in bond length and angle in the newly formed water for quite some time.
For guanine triplet (guanine-OH quartet), we continued the QM/MM simulation for 1.5ps (about times the time frame needed to see hydrogen abstraction in singlet) but there was no sign of  any significant H-OH attraction between the radical OH and the amino group guanine hydrogen as was expected from spin consideration.

\begin{figure}
\begin{center}
\includegraphics[width=1.0\linewidth]{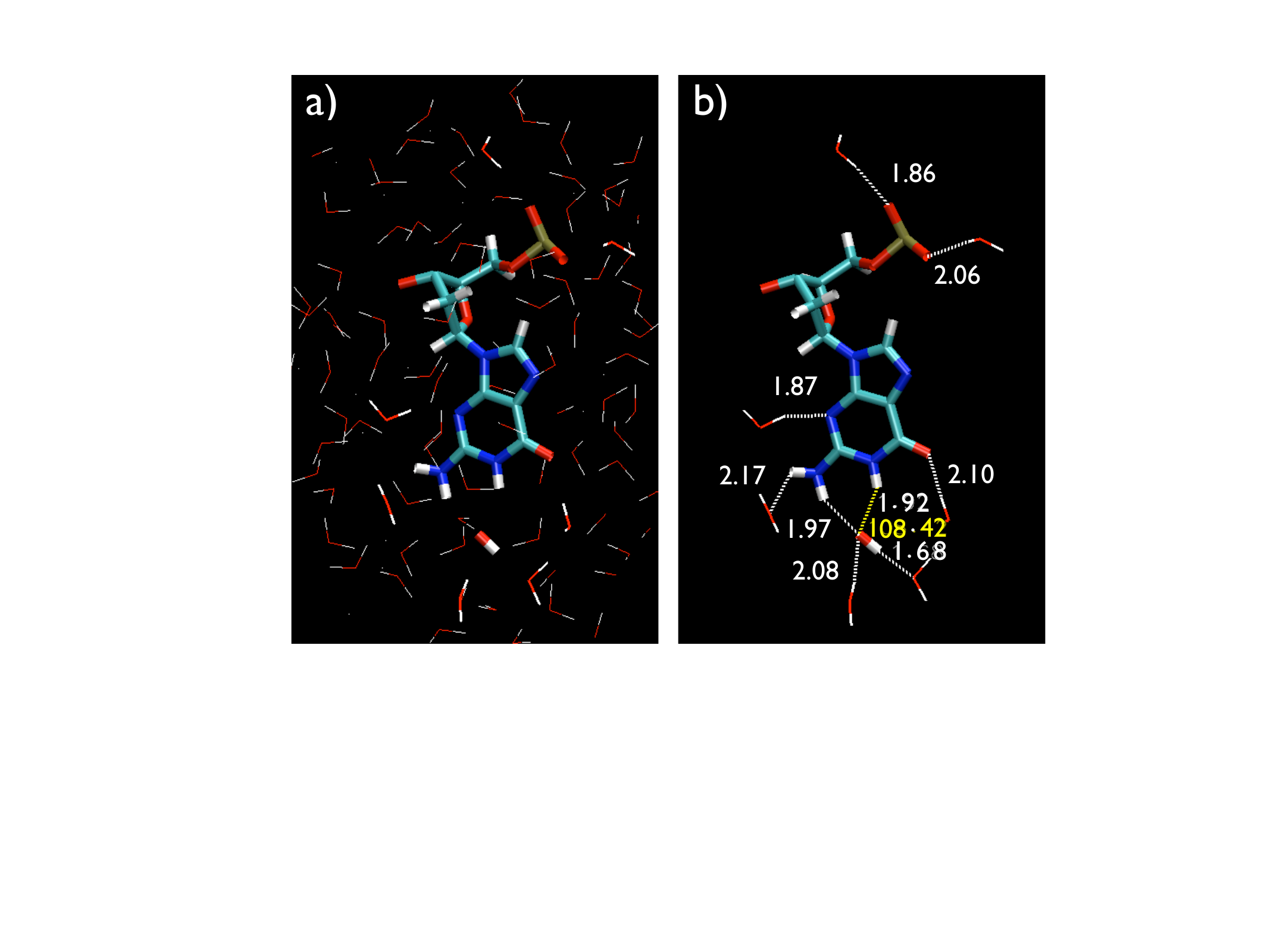}\\ 
\noindent
\caption{
A smaller subset of the complete system containing a DNA residue with guanine base surrounded by about $5 \AA$ layer of water is shown in 1a. In 1b, the water molecules which are hydrogen bonded with the DNA residue and with the OH which are directly or indirectly influencing the hydrogen abstraction process are shown.
Carbon, oxygen, nitrogen, phosphorus and hydrogen atoms are shown as green, red, blue, gold and white, respectively.
}
\label{Fig1}
\end{center}\vspace{-0.5cm}
\end{figure}

\begin{figure}
\begin{center}
\includegraphics[width=1.0\linewidth]{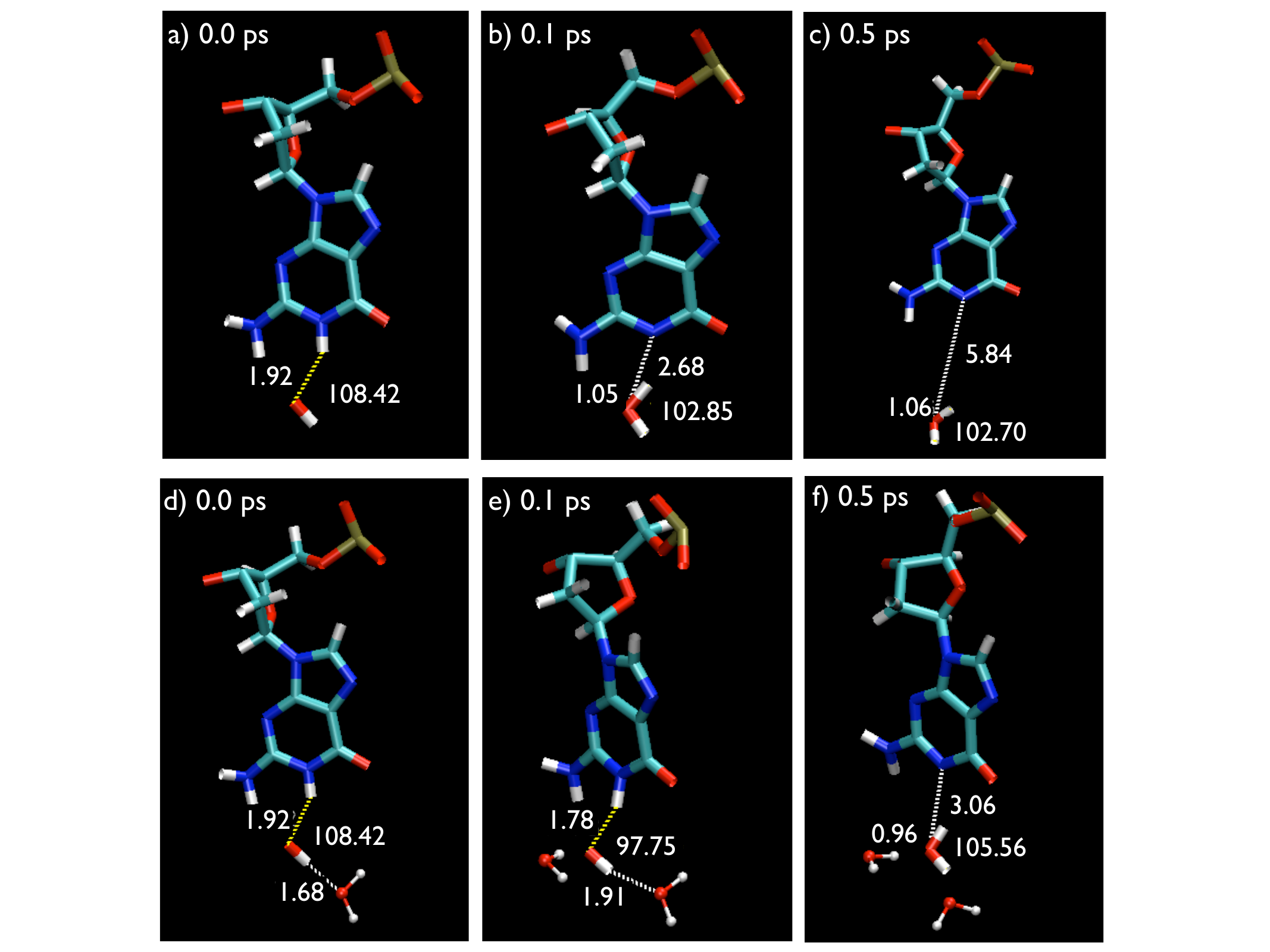}
\noindent
\caption{
Time sequence of the Hydrogen abstraction at $t=0$, $t=0.1$ ps, and $t=0.5$ ps are shown for simulations in vacuum (a,b,c) and in the presence of water (d,e,f). The Hydrogen bonds formed between water molecules in solution and the OH-guanine molecules is shown by dashed lines. The numbers in the figures denote the bond distances. The hydrogen bonds elongate the time for hydrogen abstraction.
}
\label{Fig2}
\end{center}\vspace{-0.5cm}
\end{figure}

In Fig. \ref{Fig1}(a), we show the DNA residue dissolved in water and in Fig. \ref{Fig1}(b) we show DNA residue with some of the explicit hydrogen bonding contacts with water molecule which directly or indirectly influence the hydrogen abstraction dynamics. 
The QM/MM Molecular Dynamics (MD) simulations have been performed for 1.5 ps and the MD trajectories are studied in VMD ~\cite{vmd}.
Carbon, oxygen, nitrogen, phosphorus and hydrogen atoms are shown with colors green, red, blue, gold and white, respectively.
According to our results, a dehydrogenation of the nucleotides takes place
for a system with $S_z=1/2$ (total spin-doublet).
Fig. \ref{Fig2} shows the time sequence of the hydrogen abstraction from N$_1$-site of guanine and the hydrogen bonds formation during the process that leads to the formation of a water molecule.

\begin{figure}
\begin{center}
\includegraphics[width=1.0\linewidth]{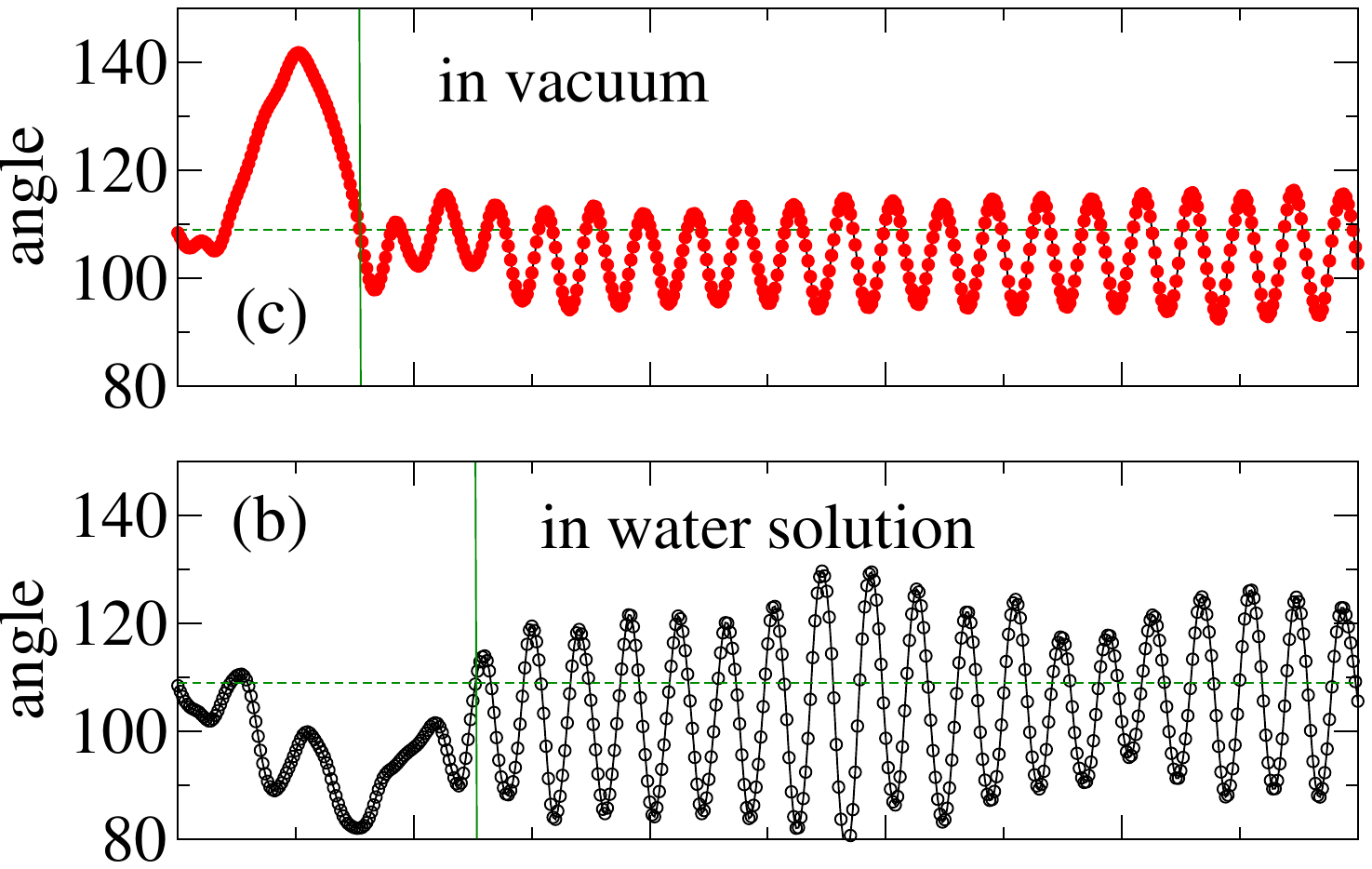}\vspace{-0.5cm}\\
\includegraphics[width=1.0\linewidth]{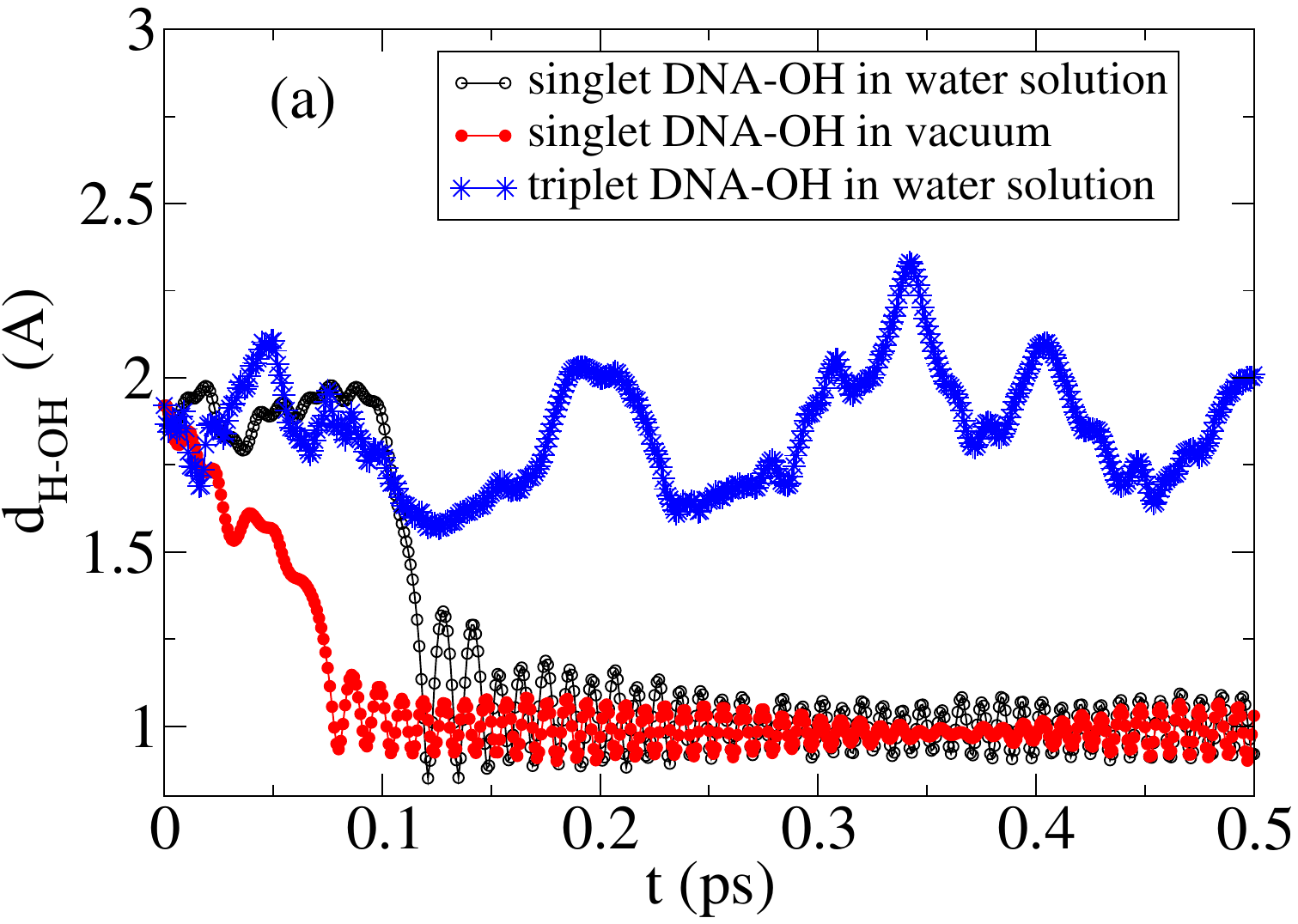}\vspace{-0.5cm}
\noindent
\caption{
In (a) the evolution of the distances (in A) from oxygen atom in the OH radical to the $H_{N1}$ in the spin singlet guanine in vacuum (filled red circles), in solution (open black circles), and in the spin triplet in solution (blue stars) are shown.
The hydrogen abstraction for singlet guanine in vacuum and solution occurs around $t\approx 0.07$ and $t\approx 0.12$ ps, respectively, and it does not occur for triplet guanine in solution.
In (b) and (c) the evolution of the angle between $H_{N1}$ and OH radical in the spin singlet guanine in vacuum (filled red circles)
and in solution (open black circles) are shown. The horizontal dashed line shows the angle 109.5 degrees. The vertical lines show the expected position of Hydrogen abstraction, consistent with the distance-time calculation shown in (a).
}
\label{Fig3}
\end{center}
\end{figure}

Fig.~\ref{Fig3}(a) shows the oxygen atom distances from the OH radical to the $H_{N1}$ in the spin singlet guanine in vacuum (filled red circles).
As it is seen the abstraction of Hydrogen occurs around $t\approx 0.07$ ps.
In solution (open black circles), however, because of the hydrogen bond network forming with water molecules in the solution and the OH-radical (see Fig.~\ref{Fig1}) the time for hydrogen abstraction is longer and it occurs at $t\approx 0.12$ ps.
This is the main finding of this study as it reveals that the reaction in solution takes place with slower rate, approximately by a factor of two, and because of the ns lifetime of OH-radicals, roughly half of them have chance of making reaction.
For comparison the spin triplet in solution (blue stars) are shown.
A drop in $H_{N1}$-OH seen in spin singlet state that is the indication of guanine dehydrogenation, disappears in case of spin triplet state where the spin-blockade effect is strong enough that it repels OH free radical.
Clearly the hydrogen abstraction does not occur for triplet guanine in solution for the time scale of the present MD simulation.
This confirms the previous calculation that the injection of $S=1$ exciton blocks the hydrogen abstraction~\cite{Abolfath2009:JPC,Abolfath2011:JPC}.
Fig.~\ref{Fig3}(b) and (c) show the evolution of the angle between $H_{N1}$ and OH radical in the spin singlet guanine in vacuum (filled red circles) and in solution (open black circles).
The horizontal dashed line shows the angle 109.5 degrees.
The vertical lines show the expected position of Hydrogen abstraction, consistent with the distance-time calculation shown in Fig.~\ref{Fig3}(a).
This confirms our hypothesis on the angular dependence of the hydrogen-abstraction of DNA molecule by free radicals, stated based on the circular symmetry of the diatomic molecule. Accordingly, the range of angles for which Hydrogen abstraction occurs is expected to be uniformly distributed in $0\leq\varphi\leq 2\pi$, but sharply distributed around $\theta=109.5$, the angle between sp3 orbitals in oxygen prior to the water-formation. After hydrogen-abstraction, as seen in Fig.~\ref{Fig3}(b) and (c), the mean angle between two hydrogens in H$_2$O molecule becomes $\theta=104.45$. Small vibrations around $\theta=104.45$ is clearly visible. These vibrations are stronger in solution than in vacuum. Here $\varphi$ and $\theta$ represent the spherical angles in a system of coordinates that the dipole-moment of OH-radical is along $\hat{z}$-axis with oxygen at the center of coordinate. This is the reflection of the circular symmetry of the diatomic molecule that strongly alters in aqueous solutions because of the hydrogen bonds between OH and neighboring water molecules.
Figs.~\ref{Fig3}(b) and (c) clearly illustrate that the alignment of the diatomic OH-radical necessary for the hydrogen-abstraction is more difficult in solution because of the thermal fluctuations of the solvent that transfers to OH-radicals through the hydrogen-bonds formation with water molecules.

\begin{figure}
\begin{center}
\includegraphics[width=1.0\linewidth]{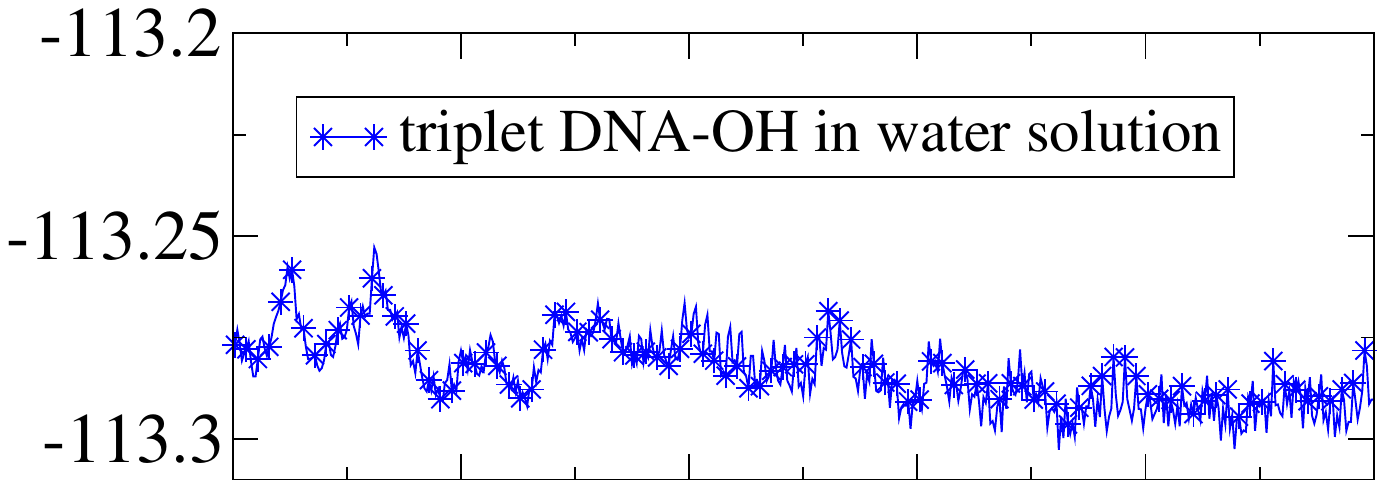}\vspace{-1.cm}\\
\includegraphics[width=1.0\linewidth]{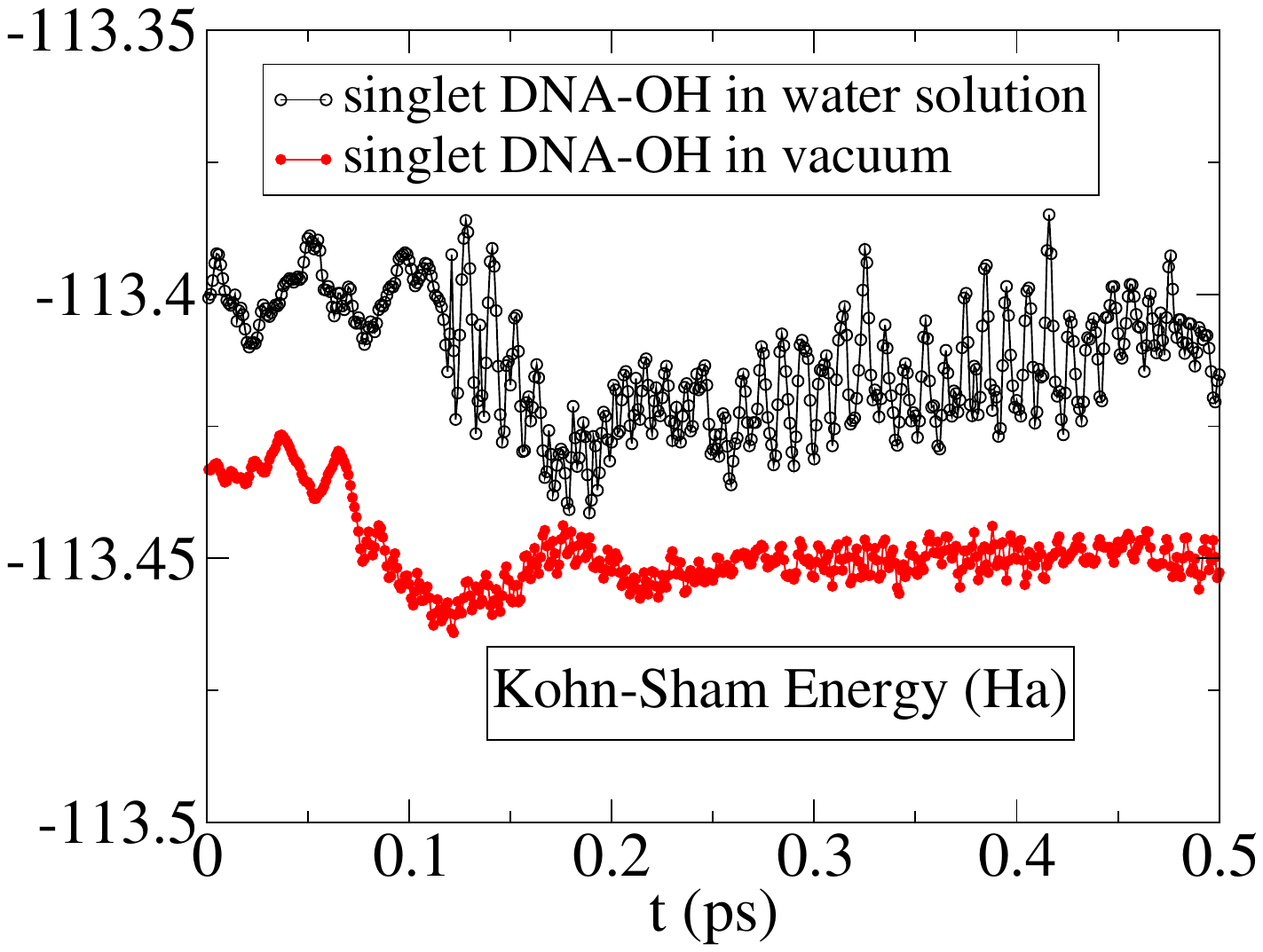}\vspace{-0.5cm}
\noindent
\caption{
The Kohn-Sham energy of the QM-part as a function of time and number of water molecules
of spin-triplet DNA (top) and spin-singlet DNA (bottom) in the presence of one OH radical.
A drop in Kohn-Sham energy is indication
of dehydrogenation of H$_{N1}$ in the guanine by OH free radical clearly seen in the spin-singlet DNA-OH
system (bottom).
The hydrogen abstraction in vacuum occurs around $t\approx 0.1$ ps.
In solution this occurs around $t\approx 0.2$ ps, approximately two times slower.
The Hydrogen-bonds coupling between water molecules and DNA residue are responsible
to strong fluctuations in Kohn-Sham energy.
}
\label{Fig4}
\end{center}
\end{figure}

Fig. \ref{Fig4} shows the Kohn-Sham energies (equivalent to potential
energy in classical molecular dynamics) of the spin singlet
and spin triplet of the guanine in the presence of the OH free radical
as a function of time, calculated by
the CPMD at T=300K corresponding to a canonical dynamics
(constant temperature ensemble).
A drop in Kohn-Sham energy, $\Delta E_{\rm KS}$, in spin singlet multiplicity is indication
of dehydrogenation of H$_{N1}$ in the guanine by OH free radical.
Our calculation shows that $\Delta E_{\rm KS}=-1.088~(-0.844)$ eV per OH-radical in solution (vacuum).
Because $|\Delta E_{\rm KS}|({\rm in-solution}) > |\Delta E_{\rm KS}|({\rm in-vacuum})$, the Hydrohen-abstraction in solution is energitically favorable, however, because of the thermal fluctuations of the water molecules that disturb the motion of OH-radical, the reaction takes place with slower rate (note that the minimum of $\Delta E_{\rm KS}$ is seen at a longer time in solution than in vacuum). Hence the network of Hydrogen bonds between water molecules and OH-radical, studied in this work, is the bottle-neck of the Hydrogen-abstraction in solution.
Finally, in the quartet spin configuration the repulsive exchange interaction, blocks the exchange of the hydrogen and hence the chemical reaction.

\begin{figure}
\begin{center}
\includegraphics[width=1.0\linewidth]{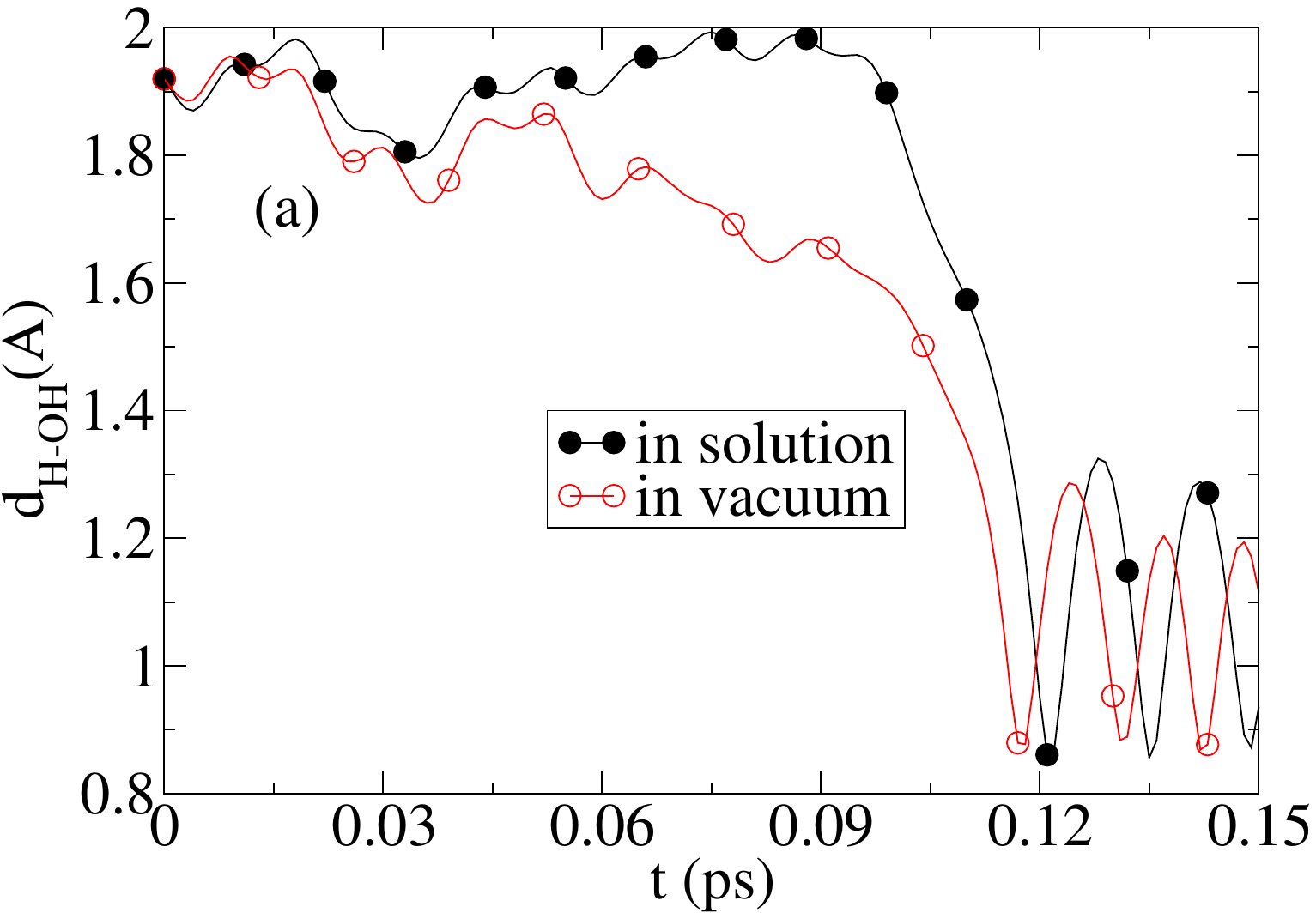}\vspace{-0.2cm}\\
\includegraphics[width=1.0\linewidth]{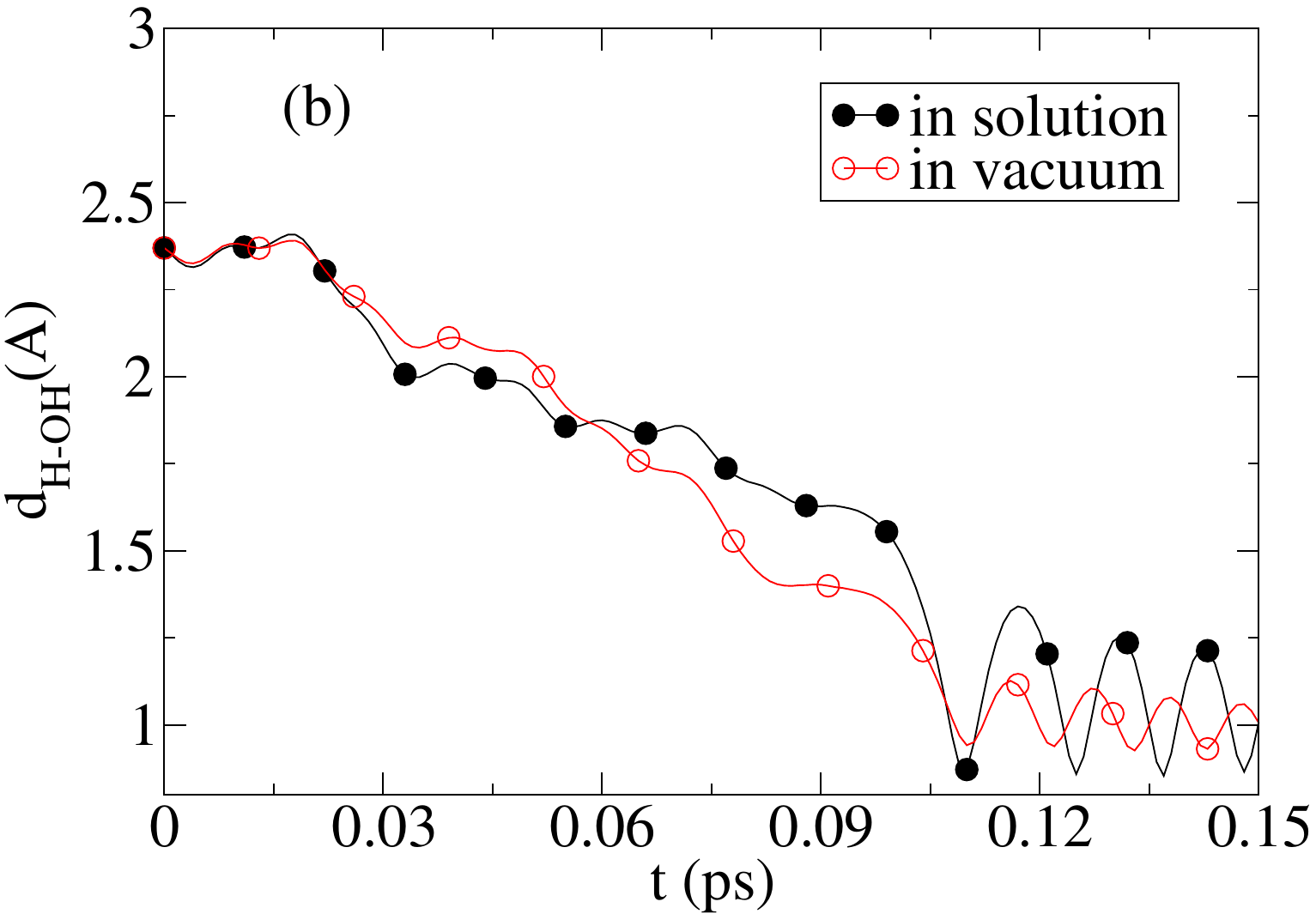}\vspace{-0.2cm}\\
\includegraphics[width=1.0\linewidth]{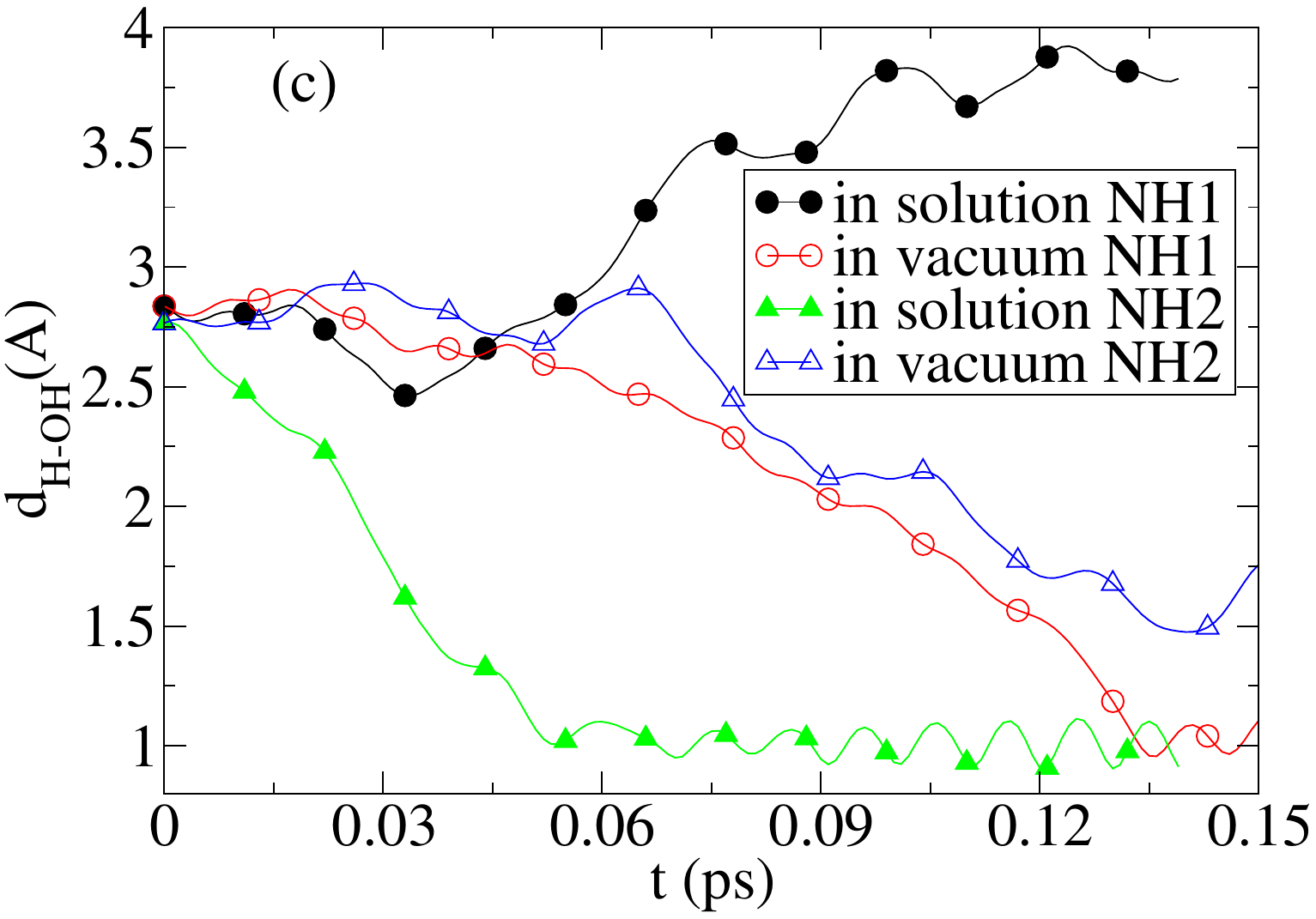}\vspace{-0.5cm}
\noindent
\caption{
The time evolution of $d_{\rm H-OH}$ as a function of initial distances $d_{\rm H-OH}=1.9,~2.3,~2.8\AA$.
In (a) and (b) $d_{\rm H-OH}$ represents the the distance between OH and $N_{H1}$ as the hydrogen abstraction occurs from $N_{H1}$. In (c) hydrogen abstraction occurs from $N_{H1}$ and $N_{H2}$ in vacuum and solution respectively.
}
\label{Fig5}
\end{center}
\end{figure}

In Figs.~\ref{Fig5} the time evolution of $d_{\rm H-OH}$ as a function of initial distance between OH and guanine in solution and in vacuum are shown. The initial OH distance varies along a line to keep OH-$N_{H1}$ and OH-$N_{H2}$ equidistance. This is a line perpendicular to a triangle edge that connects $N_{H1}$ and $N_{H2}$ as shown in Fig.~\ref{Fig1}(b).
The initial distances in Figs.~\ref{Fig5} are $d_{\rm H-OH}=1.9,~2.3,~2.8\AA$ from (a) to (c).
Comparing Figs.~\ref{Fig5}(a) and (b), we observe that the increase of initial distance from $d_{\rm H-OH}=1.9$ to 2.3$\AA$ does not alter the final product. Moreover the reaction time in both solution and vacuum appears not to be sensitive to the change of $d_{\rm H-OH}$ within this range. However, increasing initial distance to $d_{\rm H-OH}=2.8\AA$ lead to a dramatically different final state in solution where the abstraction of $N_{H2}$ is seen.
This is mainly due to effect of solution such that OH-radical gains translational and rotational energy from water molecules and crosses through transition state of OH-$N_{H2}$.

In summary, we have employed a multiscale Quantum-Mechanical-Molecular-Mechanical (QM/MM) simulation based on GROMACS-CPMD QM/MM protocol to study the hydrogen abstraction by OH-radicals from a guanine-based DNA residue solvated in explicit TIP3P water. We found that the presence of water molecules slows down the process of hydrogen abstraction by OH radicals as the latter's thermal fluctuation is restricted by its hydrogen bonding with solvent water molecules. To understand the effect of arbitrary separation of the OH radical from target hydrogen on hydrogen abstraction, we repeated the simulation on two other H-OH distances. For arbitrary separations, the  NH$_2$ hydrogens also act as potential targets for hydrogen abstraction as the thermal diffusion of OH plays a significant role in the hydrogen abstraction dynamics of OH. The present simulation, performed on a larger system of guanine attached to a deoxyribose and phosphate, exhibits that hydrogen abstraction occurs only when the guanine-H system is in its lowest spin doublet state; hydrogen abstraction is blocked for guanine-OH spin quartet. A similar effect of spin coupling on hydrogen abstraction obtained earlier with only guanine in vacuum. Present results, while complement previous calculations on smaller system in vacuum, demonstrate that such multiscale simulation, employed for the first time to address hydrogen abstraction from DNA residue, could provide a basis to address the bigger question of DNA damage by radicals in presence of solution.

We gratefully acknowledge helpful discussion and inspiring ideas with Dr. David Chen, and Dr. Michael Story.
PKB acknowledges financial support from MS-INBRE/NIH grant \#P20RR016476 and RIMI/NIH grant \#P20MD002725.





\begin{thebibliography}{99}

\bibitem{EricHall:book}
Hall, E. J. {\em Radiobiology for the Radiologist},
(Lippincott Williams \& Wilkins, Baltimore, Fifth Edition, 2000);
Becker, D.; Adhikary, A.; and Sevilla, M.; in {\em Charge Migration in DNA},
Ed. Chakraborty, T.; (Springer-Verlag, Berlin, 2007).

\bibitem{Aydogan2008:RR}
Nikjoo, H.; and Charlton, D. E.; Calculation of range and distributions of
damage to DNA by high- and low-LET radiations. In Radiation Damage
to DNA: Structure/Function Relationships at Early Times (A. F. Fuciarelli
and J. D. Zimbrick, Eds.). Battelle Press, Columbus, OH, 1995;
Nikjoo, H.; O'Neill, P.; Terrissol, M.; and Goodhead, D.T.; Int. J. Radiat. Biol. 1994; {\bf 66}, 453;
Semenenko, V. A.; and Stewart, R. D.; Radiat. Res. 2005, {\bf 164}, 180; {\em ibid.} 2005, {\bf 164}, 194;
Aydogan, B. ; Bolch, W. E.; Swarts, S. G.; Turner, J. E.; and Marshall, D. T.; Radiat. Res. 2008, {\bf 169} 223.

\bibitem{Sies1993:EJB}
Sies, H.; Europ. J. Biochem. 1993, {\bf 215}, 213.

\bibitem{Balasubramanian1998:PNAS}
Balasubramanian, B.; Pogozelski, W. K.; Tullius, T. D.;
Proc. Natl. Acad. Sci. {\bf 95}, 9738 (1998).

\bibitem{Tullius2005:COCB}
Tullius, T. D. ; Greenbaum, J. A.; Curr. Opin. Chem. Biol.  2005, {\bf 9}, 127.

\bibitem{Pogozelski1998:CR}
Pogozelski, W. K.; Tullius, T. D.; Chem. Rev.  1998, {\bf 98}, 1089.

\bibitem{Burrows1998:CR}
Burrows, C. J.; Muller, J. G.; Chem. Rev. 1998, {\bf 98}, 1109;
Cadet, J.; {\em et al.}, Mutation Res. 1999; {\bf 424}, 9.

\bibitem{Chatgilialoglu2011:CRT}
Chatgilialoglu, C.; D'Angelantonio, M.; Kciuk, G.; Bobrowski, K.; Chem Res Toxicol. 2011, {\bf 24}, 2200.

\bibitem{Herzberg:Book}
Herzberg, G.; {\em The Spectra and Structures of Simple Free Radicals} (Dover Publications Inc. New York, 1971).

\bibitem{Miaskiewicz1994:JACS}
Miaskiewicz, K.; Osman, R.; J. Am. Chem. Soc. 1994; {\bf 116}, 232.

\bibitem{Mundy2002:JPC}
Mundy, C. J.; Colvin,  M. E.; Quong, A. A.;
J. Phys. Chem. A. 2002, {\bf 106} 10063;
Wu, Y.; Mundy, C. J.; Colvin, M. E.; Car, R.;
J. Phys. Chem. A. 2004, {\bf 108} 2922.

\bibitem{Close2008:JPC}
Close, D. M.; Ohman, K. T.; J. Phys. Chem. A 2008; {\bf 112}, 11207.

\bibitem{Abolfath2009:JPC}
Abolfath, R. M.; J. Phys. Chem. B, 2009; {\bf 113}, 6938;
Abolfath, R. M.; Brabec, T.; J. Compt. Chem. 2010; 31, 2601.

\bibitem{Abolfath2011:JPC}
Abolfath, R. M., van Duin, A. C. T.; Brabec, T.; Phys. J. Chem. A 2011; 115, 11045. Online movies and animations are available at http://qmsimulator.wordpress.com/

\bibitem{Kumar2011:JPCB}
Kumar, A.; Pottiboyina, V.; Sevilla, M. D.;  J. Phys. Chem B, 2011, {\bf 115}, 15129.

\bibitem{CPMD}
Hutter, J.; Ballone, P.; Bernasconi, M.; Focher,  P.; Fois,  E.;
Goedecker, S.; Parrinello, M.; Tuckerman, M. E.; {\em CPMD code, version 3.13},
MPI fuer Festkoerperforschung, Stuttgart IBM Zurich Research Laboratory,
1990-2008.

\bibitem{wl}
Warshel, A.; Levitt, M.; J. Mol. Biol., 1976; {\bf 103}, 227.

\bibitem{singh}
Singh, U. C.; Kollman, P. A.; J. Comput. Chem., 1986; {\bf 7}, 718.

\bibitem{ego}
Eichinger, M.; Tavan, P.; Hutter, J.; Parrinello, M.; J. Chem. Phys., 1999; {\bf 110}, 10452.

\bibitem{das}
Das, D.; Eurenius, K. P.; Billings, E. M.; Sherwood, P.; Chatfield, D. C.; Hodoscek, M.; Brooks, B. R.; J. Chem. Phys., 2002; {\bf 117},  10534.

\bibitem{laio}
Laio, A., VandeVondele, J.; Rothilsberger, U.; J. Chem. Phys., 2002; {\bf 116}, 6941.

\bibitem{Biswas2005:JCP}
Biswas, P. K.; Gogonea, V.; J. Chem. Phys.,  2005; {\bf 123}, 1.

\bibitem{Yan2008:CPL}
Yan, Y.; Krishnan, G. M.; K\"uhn, O.; Chem. Phys. Lett. 2008; {\bf 464} 230.

\bibitem{opls}
Jorgensen, W. L.; Tirado-Rives, J.; J. Am. Chem. Soc., 1988; {\bf 110}, 1657.

\bibitem{opls-liq}
Jorgensen, W. L.; Maxwell, D. S.; Tirado-Rives, J.; J. Am. Chem. Soc., 1996; {\bf 118}, 11225.

\bibitem{gromacs}
Berendsen, H. J. C.; van der Spoel, D.; van Drunen, R.; Comp. Phys. Comm. {\bf 91}, 43 (1995);
Lindahl, E.; Hess, B.; van der Spoel,  D.; J. Mol. Model. {\bf 7}, 306 (2001);
Van Der Spoel, D.; Lindahl, E.; Hess, B.; Groenhof, G.; Mark, A. E.; Berendsen, H. J. C.; J. Comput. Chem. {\bf 26} 1701 (2005);
Hess, B.; van der Spoel, D.; Lindahl, E.; J. Chem. Theory Comput. {\bf 4}, 435 (2008).

\bibitem{tip3p}
Mark, P.; Nilsson, L.; J. Phys. Chem. A 2001; {\bf 105}, 9954.

\bibitem{Becke1988:PRA}
Becke, A. D.;
\pra 1988; {\bf 38}, 3098.

\bibitem{Lee1988:PRB}
Lee, C.; Yang, W.; Parr, R. G.;
\prb 1988; {\bf 37}, 785.

\bibitem{Martyna1999:JCP}
Martyna, G. J.; Tuckerman, M. E.;
J. Chem. Phys. 1999; {\bf 110}, 2810.

\bibitem{vmd}
Humphrey, W.; Dalke, A.; Schulten, K.; VMD - Visual Molecular Dynamics. J; Mol. Graphics 1996; {\bf 14}, 33.


\end{thebibliography}
\end{document}